%% file: main.tex
\documentclass{article}

\usepackage[english]{babel}
\usepackage[letterpaper,top=2cm,bottom=2cm,left=3cm,right=3cm,marginparwidth=1.75cm]{geometry}
\usepackage{amsmath, booktabs, makecell}
\usepackage{graphicx}
\usepackage[colorlinks=true, allcolors=blue]{hyperref}
\usepackage[backend=biber, style=nature, maxnames=4]{biblatex} 
\usepackage{csquotes}
\usepackage{authblk} 
\usepackage{chronology}
\usepackage[font=small,labelfont=bf]{caption} 
\usepackage{marvosym}

\newcommand{\ren}[1]{\textcolor{black}{#1}}

\providecommand{\keywords}[1]
{
  \small	
  \textbf{\textit{Keywords---}} #1
}

\title{Energy Justice and Equity: A Review of Definitions, Measures, and Practice in Policy, Planning, and Operations}

\author[1]{Weihang Ren}
\author[1]{Yongpei Guan}
\author[2,*]{Feng Qiu}
\author[2]{Todd Levin}
\author[3]{Miguel Heleno}
\affil[1]{\normalsize Department of Industrial and Systems Engineering, University of Florida, Gainesville, FL, USA}
\affil[2]{\normalsize Energy Systems Division, Argonne National Laboratory, Lemont, IL, USA}
\affil[*]{fqiu@anl.gov}
\affil[3]{Lawrence Berkeley National Lab}

\date{}

\begin{document}
\maketitle

\begin{abstract}
    Energy justice, at the intersection of energy and societal ethics, studies the origins, quantification, and resolution of persistent and potential inequities within the energy sector, serving as a foundational pillar for societal harmony.
    In this review, we overview the historical and modern definitions of energy equity and frameworks of energy justice.
    We highlight the tools adopted to measure equity in the energy context, unveiling multifaceted inequities that permeate global energy landscapes.
    We discuss the limitations of prevalent metrics such as the Gini coefficient and Generalized Entropy Indices in the evaluation of energy justice concerns.
    Finally, we analyze publications that examined current practices and proposed improving methods toward a more equitable energy market for society from policy, planning, and operation perspectives.
\end{abstract}


\bigskip

\input{body}


\printbibliography 

\end{document}

%% file: body.tex
\begin{figure}[htbp]
    \centering
    \includegraphics[width=0.7\textwidth]{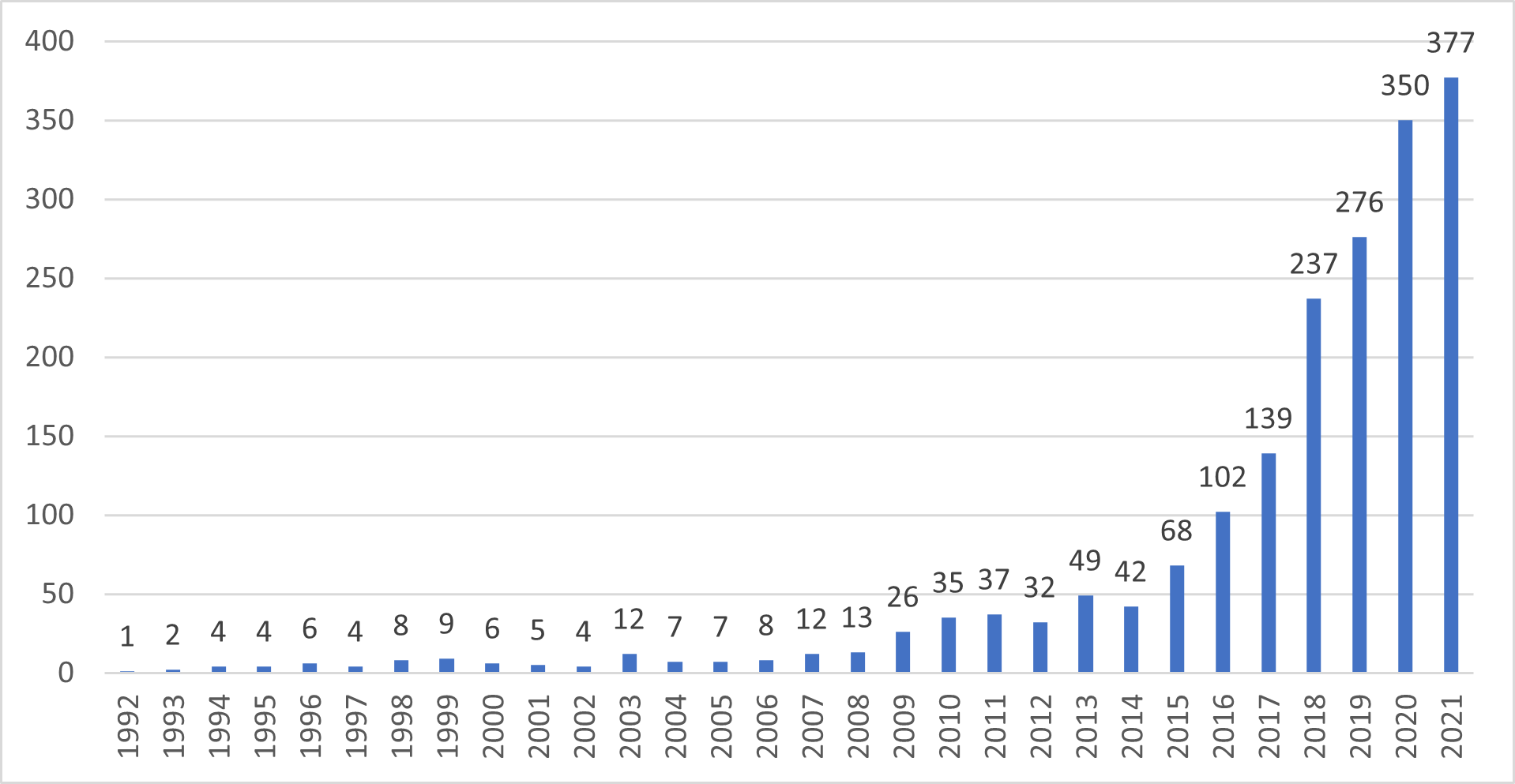}
    \caption{Annual number of publications on ``energy justice'' from 1992 to 2021, counted by \cite{qian2023survey}.}
    \label{fig:count}
\end{figure}

In the last three decades, energy justice has emerged as a subject of considerable discourse among research communities and policymakers. 
Energy justice refers to the fair distribution of benefits and burdens associated with energy production, distribution, and consumption. It encompasses the idea that all individuals, regardless of socioeconomic status, race, or geographic location, should have equal access to the benefits of clean and affordable energy, as well as a fair share of the costs and risks associated with energy development. The understanding, definition, measurements, and priorities on energy justice are also evolving, especially in recent years as renewable penetration and electric vehicle adaption surge, power market reform continues, and new business models emerge. These latest developments in the energy sector, plus significant disruptions by recent pandemics and extreme weather events, have created more challenges for energy justice. In a significant stride towards this goal, the 2021 Biden administration's {\it Justice40 Initiative} \cite{order8614008} aims to deliver at least $40\%$ of the overall benefits of relevant federal investments in climate and clean energy to disadvantaged communities. Notably, the number of publications on energy justice during 2019-2021 (1003) exceeds the total number of publications in the previous three decades (879) \cite{qian2023survey}, as shown in Figure~\ref{fig:count}. To better understand and address energy justice in this complex and dynamic environment, in this review, we delve into the historical evolution of energy justice and equity definitions and summarize prevalent metrics employed to reveal the various forms of energy inequities around the world, highlighting efforts in policies, grid planning, and operations that dedicated to ushering justice into energy systems.


\section{Energy Justice and Equity in Government Definitions}

Energy justice and energy equity are crucial concepts affecting society in many ways. Before delving into the broad academic research on them, we start with the government definitions that reflect the comprehensive ideas behind these terms.

\subsection{Energy Justice}\label{def-justice}

As energy is an indispensable branch of the environment, there was no formal government definition of energy justice alone in the 20th century, instead, the energy justice concept was covered under the umbrella of {\it environmental justice}. The latter was provided in 1994 during President Clinton's administration and defined as ``identifying and addressing, as appropriate, disproportionately high and adverse human health or environmental effects of its programs, policies, and activities on minority populations and low-income populations'' \cite{clinton1994executive}. Based on this definition, energy justice could be interpreted as applying justice principles to the health and environmental consequences of energy programs, energy policies, and energy activities on minority and low-income populations. 

As renewable energy increasingly penetrates the power grid, the energy price gets more volatile, among others, and accordingly, disproportionately distributed benefits from renewable energy broaden the coverage of energy justice. The economic consequence has thus stood out in the government definitions.
In 2019, the US Department of Energy (DOE) defined {\it energy justice} as ``the goal of achieving equity in both the social and economic participation in the energy system, while also remediating social, economic, and health burdens on those disproportionately harmed by the energy system'' \cite{energyjusticedoe}. As compared to the 1994 interpretation, this definition specifically highlights a new dimension of achieving equity (with the definition below in Section~\ref{def-equity}) in both the social and economic participation in the energy system, besides health and environmental concerns. 

Most recently, along this direction, a guideline regarding a quantitative measure has been introduced in 2021 by President Biden's Administration to characterize justice. It is called {\it Justice40 Initiative}, which guides certain Federal investments intending to direct 40 percent of the overall benefits in climate and clean energy to disadvantaged communities \cite{order8614008}. The ``disadvantaged communities'' are specifically identified by DOE through 36 indicators covering fossil dependence, energy burden, environmental and climate hazards, and vulnerability (socioeconomic, housing burden, transportation burdens, etc.) \cite{doe40}. This guideline delineates a quantifiable goal and adds more dimensions by not limiting the subject to specific injustice matters. 

\subsection{Energy Equity}\label{def-equity}

As mentioned in Section~\ref{def-justice}, energy equity is a term associated with energy justice. While energy justice sets goals of justice in energy systems as mentioned above, energy equity provides means to achieve these goals, such as through energy accessibility, participation in energy programs, and fair treatment of all individuals. 

The definition of energy equity evolves from a simple concept to specified implementable policies. The earliest government definition of energy equity dates back to 2015 when the United Nations (UN) advocated equality and non-discrimination in energy usage in its Sustainable Development Goals (SDGs) and envisaged a world where ``the needs of the most vulnerable are met.'' Specifically, SDG 7 aimed to provide all people with affordable, reliable, sustainable, and modern energy by 2030 \cite{desa2016transforming}. It defines {\it energy equity} as unlocking access to energy for everyone to realize the goal of energy justice.

To achieve this, the main focus should be empowering people who are most vulnerable to access energy. Accordingly, DOE launched the Equity in Energy initiative in 2020. It aims to ``expand the inclusion and participation of individuals in underserved communities, such as minorities, women, veterans, and formerly incarcerated persons, in all the programs of the DOE and the private energy sector \cite{equityinenergy}.'' This initiative focuses on the underserved communities out of all populations, which are usually the most vulnerable groups, gathering the resources where they are needed the most.

Lately, the ``equity" and ``underserved communities" have been specified by the US government in 2021, to enrich the coverage of energy equity to better carry out policies. The government defined ``equity" as ``the consistent and systematic fair, just, and impartial treatment of all individuals, including individuals who belong to underserved communities that have been denied such treatment, such as Black, Latino, and Indigenous and Native American persons, Asian Americans and Pacific Islanders and other persons of color; members of religious minorities; lesbian, gay, bisexual, transgender, and queer (LGBTQ+) persons; persons with disabilities; persons who live in rural areas; and persons otherwise adversely affected by persistent poverty or inequality." It further defined ``underserved communities" as ``populations sharing a particular characteristic, as well as geographic communities, that have been systematically denied a full opportunity to participate in aspects of economic, social, and civic life, as exemplified by the list in the preceding definition of `equity'"\cite{order8613985}. This definition provides a detailed guideline for society to actually implement energy equity policies to achieve justice. In addition, this state-of-the-art categorization of the underserved communities highlights the modern focus of the energy equity programs.

At the end of this section, we summarize the timeline and evolution process of energy justice and energy equity government definitions in Table~\ref{tab:time}.

\begin{table}[ht]
\caption{Summary of government definitions on energy justice and energy equity}\label{tab:time}
{\small
\begin{tabular}{lll p{0.24\linewidth} p{0.24\linewidth}}\toprule
Concept  & Institute   & Time & Subject & Target \\ \midrule
Environmental   Justice \cite{clinton1994executive} & White House & 1994 & Health or environmental effects & Minority and low-income populations \\
Energy Equality \cite{desa2016transforming} & UN  & 2015 & Access to energy  & All individuals \\
Energy Justice \cite{energyjusticedoe} & DOE  & 2019 & Participation in the energy system; burdens & Those disproportionately harmed by the energy system \\
Equity in Energy \cite{equityinenergy} & DOE  & 2020 & Inclusion and participation in energy programs & Underserved communities  \\
Equity \cite{order8613985} & White House & 2021 & Consistent and systematic treatment  & All individuals; underserved communities \\
Justice40 \cite{order8614008} & White House & 2021 & Federal investments  & Disadvantaged communities \\ \bottomrule
\end{tabular}
}
\end{table}

\section{Energy Justice and Equity in Literature}
The energy fairness issues have been studied in the academic domain earlier than the government definitions were provided. As compared to high-level conceptual governmental definitions, energy justice, and energy equity in the academic domain focus on establishing detailed analytical frameworks and quantitative methods on injustice, unfair, or disproportionate issues. 
More specifically, the lens of energy justice in literature analyzes justice issues from three or more dimensions emphasizing various aspects of justice, such as the distribution of resources, the decision-making procedure, and the recognition of injustice. The concept of energy equity is further specified by measuring the severity of the sufferings resulting from energy injustice, covering common types of unfair treatment in energy-related areas.
It is worthwhile mentioning here that different concepts, such as energy democracy, have been proposed in the literature to promote individual participation in decision-making processes relating to energy systems,
and eventually are included in the energy justice framework \cite{droubi2022critical}.

\subsection{Energy Justice}
As a conceptual notion derived from justice, energy justice in literature can be summarized as a framework for analyzing the fair and impartial accessibility and opportunity for all individuals involved in energy systems. This analytical framework mainly consists of three core tenets: distributional justice, procedural justice, and recognition justice \cite{jenkins2016energy, carley2020justice, van2021energy}. 

Distributional justice initially addressed the locational disparities of benefits or burdens related to energy~\cite{jenkins2016energy}. The purpose of distributional justice is to identify and avoid situations where some populations carry a disproportionate share of the burdens or lack access to benefits. This concept has been extended in~\cite{carley2020justice} from across geographic locations to across populations, because the locational difference can be seen as a character within a community, among other characteristics. It is worth noting that, while the unfair distributions across locations and economic statuses have been studied, the unequal treatments across gender, race, and other characteristics still require more attention \cite{cannon2021gender}.

Procedural justice is concerned with who has access to energy decision-making processes. It aims for equitable energy procedures by encouraging the inclusiveness of persons involved at all levels. Three mechanisms to facilitate inclusion for procedural justice were proposed, namely, local knowledge mobilization, full information disclosure, and equal institutional representation \cite{jenkins2016energy}. Especially, overlooking Indigenous knowledge in energy decision-making can be detrimental to the local community \cite{mccauley2015assessing}.

Recognition justice calls for efforts to identify individuals affected by energy injustices. While some publications include it as part of procedural justice given its fair representation feature \cite{van2021energy}, many take it as a separate tenet because it recognized energy injustice from more perspectives than the coverage of procedural justice, such as historical energy inequalities \cite{carley2020justice}. Failure to recognize injustice situations postpones the realization of energy justice. Three main categories of misrecognition are cultural domination, non-recognition, and disrespect \cite{fraser1999social}.

Some researchers add a fourth dimension to the framework: restorative justice, which describes the remediation of a perceived energy injustice \cite{van2021energy}. As a concept primarily used in criminal law, restorative justice can be applied to each phase of the energy life-cycle to repair the harm caused by injustice \cite{heffron2017concept}.

By utilizing this framework, researchers can analyze and uncover existing energy injustice-related issues. For example, one study applied the framework to explore urban residential energy consumption for heating and discovered the significance of spatial, racial, and socioeconomic factors in the unjust distribution of energy efficiency in Detroit, Michigan \cite{reames2016targeting}. 
In another study, the framework was used to survey and categorize energy equity metrics (with the details to be described in the following section), such that the focus of each metric in terms of energy justice dimensions is identified, which helps facilitate interactions among metric users \cite{barlow2022advancing}. Moreover, researchers have utilized energy justice to investigate specific issues in energy policy, planning, and operations, including energy transitions \cite{baker2021marginalized, carley2020justice}, community energy initiatives \cite{van2021energy}, gas policy and planning \cite{cotton2017fair}, power plant pollution \cite{frischkorn2020power}, energy storage \cite{tarekegne2021energy}, and electricity tariff designs \cite{khan2021electricity}. Finally, as a by-product, researchers surveyed energy programs across the US and identified 253 energy justice programs in total, with 83\% of them being run by nonprofit organizations, indicating less effective actions from governments (less than 17\%) \cite{carley2021analysis}. 

\subsection{Energy Equity} \label{sec:def-lit-equity}
Within the framework of energy justice, ``energy equity'' serves as a comprehensive term encompassing various forms of inequities experienced by individuals and communities as a whole. To gauge the extent of inequity within energy-related situations, ``energy equity'' can be further divided into distinct categories, such as energy poverty, energy burden, and energy insecurity. Some publications have previously used these notions interchangeably \cite{reames2016targeting, brown2020high, goldstein2022racial}, but recent studies have increasingly emphasized their unique angles in characterizing situations that challenge the notion of energy equity. 

Tarekegne et al. \cite{tarekegne2021review} summarized four primary categories within the realm of energy inequity: energy poverty, energy burden, energy insecurity, and energy vulnerability. While the boundaries of each category do not have a universal consensus in research \cite{qin2022impact}, a rigorous delineation of these concepts holds significant importance. Such a classification not only standardizes research within the field of energy justice but also facilitates the identification of those who are disproportionately affected. By dissecting and distinguishing these categories based on demographic and energy-related indicators \cite{hernandez2015sacrifice}, this approach enhances our understanding of energy inequities.

In this section, we aim to categorize and present the prevalent definitions, status quo, and challenges associated with these equity angles. It is important to note that our focus here is on these inequity terms within the context of households, they should not be confused with the same words in other contexts, such as ``energy burden" in the context of energy generation systems \cite{park2021water} or ``energy insecurity" for a nation \cite{mayer2022fossil}.

\subsubsection{Energy poverty}

Energy poverty, defined as inadequate access to modern energy services, stems from various causes, including a lack of energy infrastructure and utility disconnections due to the inability to pay the energy bills \cite{goldstein2022racial}. For instance, in Africa, the persistence of energy poverty is exacerbated by the high costs associated with extending power grids, while in South American urban communities, where the grid exists, unpaid utility bills pose significant challenges \cite{baker2021marginalized}. Energy poverty extends beyond electricity; it encompasses inadequate access to cooking and heating fuels, transportation gas, and other essential daily energy needs, making its eradication a complex endeavor. 

The consequences of energy poverty are far-reaching, including social exclusion, material deprivation, and poor health resulting from the lack of access to essential lighting, cooling, or heating services \cite{hilbert2016turn}. Additionally, gender and ethnicity disparities play a substantial role in perpetuating energy poverty, further exacerbating the imbalance in access to energy resources \cite{rosenberg2020evidence, luke2021powering, NGARAVA2022112755}. As an intersection of poverty and the energy system, energy poverty contradicts the fundamental goal of energy equity, resulting in unequal access to vital energy resources. 

Alarmingly, as of 2021, approximately 770 million people worldwide still lack access to electricity, with the majority residing in Africa and developing Asian countries \cite{iea2021world}. The Coronavirus Disease 2019 (COVID-19) disrupted the steady progress towards universal access to electricity and clean cooking. In response, researchers and policymakers have been actively addressing these challenges through initiatives such as rural electrification \cite{AKBAS2022111935} and revising energy pricing strategies \cite{ansarin2022review} to promote greater accessibility among marginalized communities.

\subsubsection{Energy insecurity}

Energy insecurity is defined as the inability to adequately meet household energy needs \cite{tarekegne2021review}. It is worth clarifying that energy insecurity is a term often applied in the studies for developed countries, in contrast to ``energy poverty'' which is more commonly used in discussions of less developed regions \cite{mohr2018fuel}. Interestingly, some studies in developed countries use ``energy poverty" interchangeably with ``energy insecurity" \cite{scheier2022measurement}. 

The notion of energy insecurity can be traced back to ``fuel poverty.'' Boardman \cite{boardman1991fuel} first used ``fuel poverty'' to describe the homes that could not afford adequate warmth. This concept invoked rigorous studies on fuel pricing and energy efficiency in the United Kingdom (UK) and Northern Europe. Unlike ``energy poverty,'' people who suffer from fuel poverty have access to the required energy service but struggle with affordability. 


Households often experience energy insecurity issues as a result of extreme heat or cold-related events. Yoon and Hern{\'a}ndez \cite{yoon2021energy}, through a thematic analysis of US news media articles from 1980 to 2019, shed light on geographical distinction in energy insecurity risks. They found that Southern states were the predominant victims of extreme heat, whereas Midwestern and Northeastern states had more coverage of extreme cold-related issues. The authors emphasized that income and wealth inequalities within states played a significant role in shaping the disproportionate impact of extreme temperatures on various demographic groups.

However, Baker et al. \cite{baker2021perspective} warned that ethnicity disparities in energy insecurity were often more severe than the disparities based on socioeconomic levels. Memmott et al. \cite{memmott2021sociodemographic} further showed that households at or below $200\%$ of the federal poverty line, as well as Black and Hispanic households, were more likely to experience energy insecurity. 
 
\begin{figure}[htbp]
    \centering
    \includegraphics[width=0.8\textwidth]{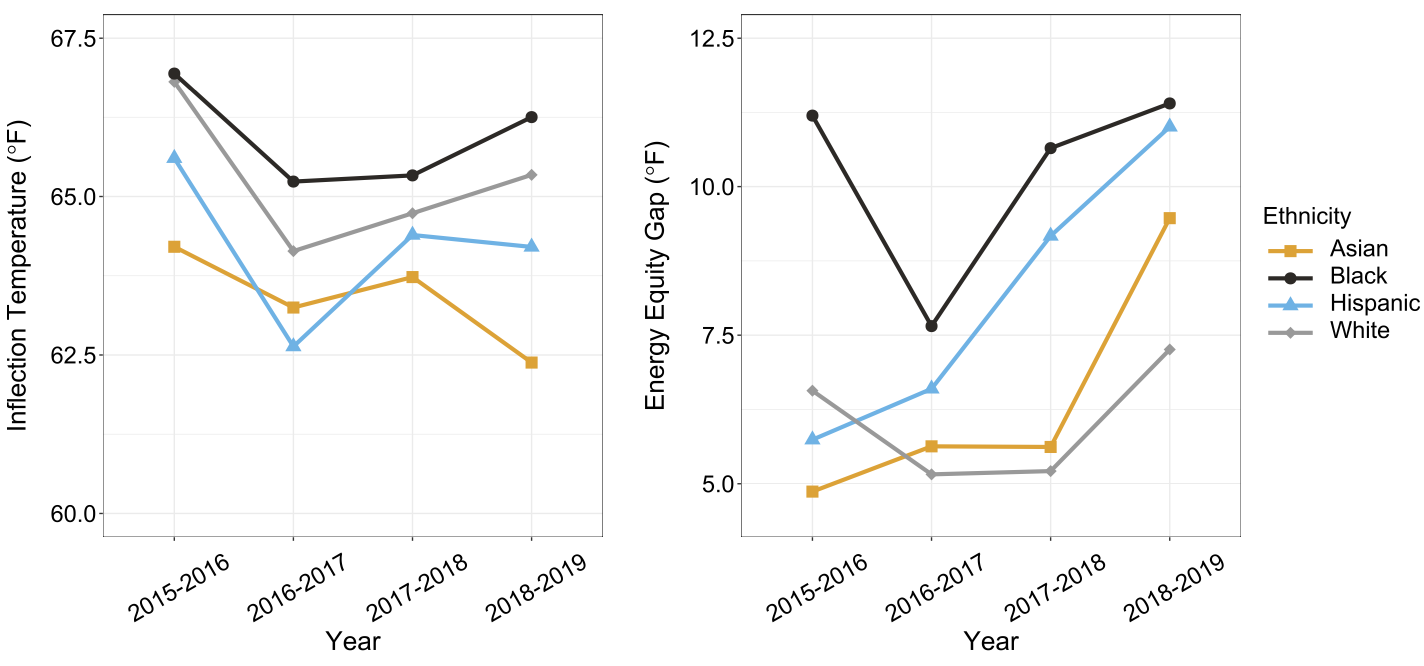}
    \caption{Inter- and Intra-group Comparison of the Inflection Temperature \cite{cong2022unveiling}. Left: Median inflection temperature shows disparities across ethnicity. Right: The energy equity gap illustrates the difference in energy consumption behavior between high and low-income groups within each group.}
    \label{fig:unveil}
\end{figure}

A critical issue highlighted by Bednar and Reames \cite{bednar2020recognition} is that, despite its prevalence, the United States has not formally recognized energy insecurity as a distinct problem from general poverty at the federal level, limiting effective responses. They argued for a shift in current metrics, which primarily focus on resource distribution, toward metrics that prioritize improving household well-being and reducing overall energy insecurity. In this regard, Cong et al. \cite{cong2022unveiling} developed a novel energy insecurity metric known as the ``energy equity gap.'' This metric is calculated based on the difference in the inflection temperatures between low and high-income groups, with inflection temperature defined as the outdoor temperature at which a household begins to use its cooling system. Figure~\ref{fig:unveil} illustrates the disparities in inflection temperatures. The figure on the left compares the median inflection temperature for each ethnicity over a span of 4 years. The figure on the right presents the "Energy Equity Gap" within each ethnicity over the same period, where each data point represents the difference in inflection temperatures between low- and high-income populations for a given ethnicity. The figure shows increasing Energy Equity Gaps for all four races from 2017 to 2019. The Black population experienced both the highest overall inflection temperature and the most significant energy equity gap during the years 2015-2019.

Recent publications have also examined the disproportionate impact of the pandemic on energy insecurity. Lou et al. \cite{lou2021inequitable} analyzed the percentage changes in hourly electricity consumption resulting from measures taken against COVID-19 and found an increasing trend in energy insecurity among ethnic-minority households and low-income households. 

\subsubsection{Energy burden}

Energy burden measures the percent of a household’s income spent to cover energy cost \cite{day2020equitable}. This financially induced definition implies low affordability of energy for low-income families. When the energy burden surpasses a certain threshold, it can push a family into a state of energy insecurity. In many studies, $6\%$ has become a widely used threshold, based on the idea that families can reasonably allocate up to $30\%$ of their income on shelter costs, with energy bills usually accounting for $20\%$ of that expenditure \cite{nysix2019understanding}. 

\begin{figure}[htbp]
    \centering
    \includegraphics[width=0.9\textwidth]{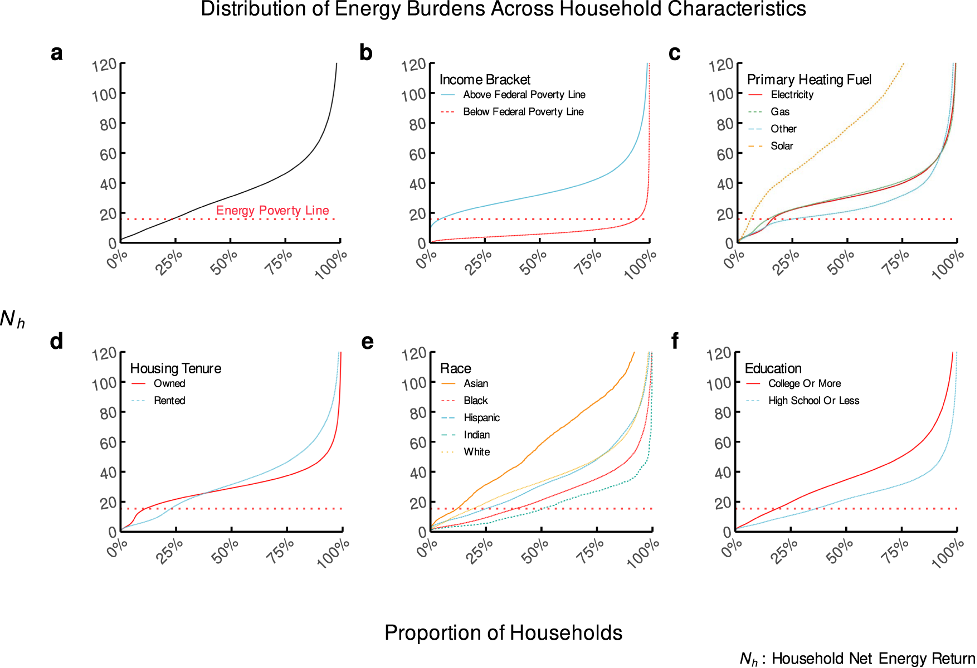}
    \caption{The Distribution of $N_h$s Across Different Household Characteristics \cite{scheier2022measurement}. Subfigure \textbf{a} shows the overall distribution. Subfigures \textbf{b}-\textbf{f} show the difference among groups of households with different characteristics.}
    \label{fig:burden}
\end{figure}

Scheier and Kittner \cite{scheier2022measurement} highlighted a lag in energy burden studies in the US due to the absence of more modern metrics. The energy burden formula results in a long tail of households with very low incomes and fails to capture zero income and negative energy cost scenarios. To address this gap, they introduced the Household Net Energy Return ($N_h$), defined as:
\begin{equation}
    N_h = (G-S)/S,
\end{equation}
where $G$ represents the gross income for a household and $S$ denotes their spending on energy. With the established threshold for energy burden at $S/G = 6\%$, the energy insecurity line for $N_h$ is approximately $16$. In practical terms, this means that a household earning less than \$$16$ for each dollar spent on energy will be considered energy insecure (or ``energy poor" in the developed country context). This is shown as the horizontal red dotted line in Figure~\ref{fig:burden}\textbf{a}, where the horizontal axis shows the proportion of households and the vertical axis shows the Net Energy Return. The black curve in Figure~\ref{fig:burden}\textbf{a} shows the distribution of $N_h$ across all populations.

It is crucial to emphasize that, despite the financial nature of the notion of energy burden, the income gap is not the only aspect that needs to be addressed. Some other important household characteristics are shown in Figure~\ref{fig:burden}c-f, including primary heating fuel type, housing type, household race, and education. As Executive Order 13985 \cite{order8613985} advocated, all underserved communities who have been denied opportunities to modern energy services are worth further attention from researchers and policy-makers. Consequently, there is a growing need to establish effective and tailored metrics to facilitate the recognition and support of energy-burdened families. 

Bird and Hern{\'a}ndez \cite{bird2012policy} estimated potential annual savings ranging from $\$4$ billion to $\$11$ billion by addressing the energy burdens for the poorest renters in the United States. To facilitate the recognition and assisting process for energy-burdened families, the U.S. DOE developed the Low-income Energy Affordability Data (LEAD) Tool \cite{ma2019low}. This tool provides census-tract-level data on low-income household populations and their associated energy use characteristics. 

\subsubsection{Energy vulnerability}

Energy vulnerability, a lesser-explored aspect of energy equity, describes the propensity of a household to suffer from a lack of adequate energy services due to an advert event or change \cite{tarekegne2021review}. Households facing severe energy vulnerability may in turn experience high energy burden, energy insecurity, or even energy poverty in the foreseeable future. This vulnerability often arises from mismatches between the supply and demand of energy, influenced by a multitude of factors. These factors include risks associated with weather and climate, which are common causes of changes in energy demand and setbacks in energy generation and transmission. Additionally, trade patterns and geopolitical concerns can cause volatile energy prices \cite{iea2021world}. 

\begin{figure}[htbp]
    \centering
    \includegraphics[width=0.9\textwidth]{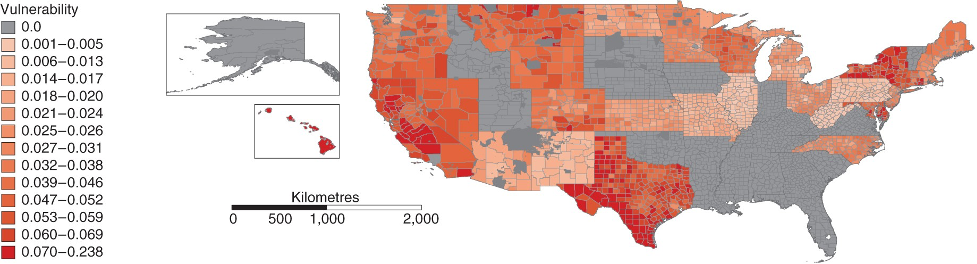}
    \caption{Vulnerability Score Map \cite{carley2018framework}. Each colored range of the vulnerability score has the same number of observations.}
    \label{fig:vulnerability}
\end{figure}

Recent studies show that the transition and development of the energy grid can affect households' energy supply unintentionally. Carley et al. \cite{carley2018framework} provided a framework for understanding energy vulnerability from energy transitions across three dimensions: exposure to risk, sensitivity to the impacts of these risks, and adaptive capacity to attenuate, cope with, or mitigate the negative effects. To quantify vulnerability, the authors proposed a Vulnerability score $V$ based on this framework, calculated as follows:
\begin{equation}
    V = \sum_{p=1}^P \left(1/I \sum_{i=1}^I E_i \left(\sum_{j=1}^J S_j \right)_i - 1/I \sum_{i=1}^I E_i \left(\sum_{k=1}^K A_k \right)_i \right),
\end{equation}
where $p$ is each policy (risk) under consideration, $E_i$ measures the exposure of type $i$ associated with a policy, $S_j$ and $A_k$ evaluate the sensitivity of type $j$ and adaptive capacity of type $k$ for a type of exposure respectively. A higher vulnerability score indicates less resilience to the risk for a community. In Figure~\ref{fig:vulnerability}, the authors demonstrated significant geographic variation in energy vulnerability by 2013. 

Numerous publications have attempted to enhance energy resilience (mitigate energy resilience) through initiatives related to housing and energy supply infrastructure \cite{maxim2022anticipating}, vulnerability indices \cite{gatto2020energy}, and trading policies \cite{pan2022design}. While efforts to improve energy resilience benefit those directly involved, it is important to acknowledge that the unequal distribution of energy vulnerability across the country remains a significant concern. Factors influencing energy vulnerability are expected to exacerbate such injustices. 


Chen et al. \cite{chen2022extreme} addressed disproportionate impacts of disasters and pandemics from both micro-level (socio-demographic and household factors) and macro-level (quality of energy and infrastructure) perspectives. They discussed the applications of mathematical and statistical models to understand and predict the outcomes of disasters, further highlighting the multifaceted nature of energy vulnerability within communities.

\subsubsection{Beyond existing categories}

While these four energy equity categories provide valuable building blocks for studying energy inequity situations based on the accessibility and affordability of energy resources to households and communities, they fall short in capturing the full spectrum of energy equity issues. More perspectives must be explored to more precisely and effectively address energy justice. 

A crucial aspect often overlooked is the differentiation of energy resources and energy services. Various energy types and evolving energy services have distinct impacts on consumer behaviors and priorities. For instance, the utility bills associated with electricity used for cooking may differ significantly from those used for charging electric vehicles. These nuances are not given sufficient attention in current energy equity discussions.

Moreover, the portrayal of ``underserved communities" and ``disadvantaged communities" based solely on income, race, and location is limited. Ivanova and Middlemiss \cite{ivanova2021characterizing} highlighted that households with disabled members often earn less and consume $10\%$ less than other households in the European Union (EU), making them more susceptible to energy poverty. Baker et al. \cite{baker2021perspective} made a critical distinction between marginalized communities and low-income neighborhoods, emphasizing that marginalized populations are often underrepresented in decision-making processes related to energy justice. Raising the voices of underrepresented populations can introduce greater diversity into studies and help overcome the inertia within traditional frameworks. Pinpoint the target persons and communities who have suffered from energy injustice can enhance the recognition process and improve the effectiveness of assistance programs. 
 
Furthermore, despite the innovative scores we have reviewed earlier, there remain energy equity issues that lack a satisfying quantitative measure. Additionally, special attention should be given to the localization of energy metrics derived from foreign research. When translating energy metrics across countries, these methods often necessitate the normalization of various variables among different types of data, which can be a complex and challenging endeavor \cite{scheier2022measurement}. This highlights the need for further research and the development of more adaptable and context-specific metrics to address energy justice on a global scale.


\section{Measuring Energy Justice}

While many methods were applied to recognize and correct energy inequities, such as regression models \cite{mohr2018fuel, ivanova2021characterizing, goldstein2022racial, alola2022nexus}, net energy analysis \cite{scheier2022measurement}, along with various data mappings and visualizations \cite{sunter2019disparities, sasse2020regional, brockway2021inequitable}, there are well-established equity measuring tools from economics and network communications being widely adopted in energy equity studies. The most popular instruments are the Gini coefficient and Generalized Entropy indices. In this section, we introduce the definition of each measurement, present their applications in energy equity, and analyze the shortcomings of these tools.

\subsection{Gini Coefficient}

\subsubsection{Definition of Gini Coefficient}
Gini coefficients have been widely used in economics to estimate income inequality as an indicator of social welfare \cite{yitzhaki2013gini} thanks to its simple mathematical formulation and instinctive graphical interpretation of inequality. The Gini coefficient quantifies inequality as the difference between an ideal situation of equality where everyone has the same shares of a resource and the actual distribution of the resource. It ranges from $0$ to $1$ with value $0$ denoting the perfect equality while value $1$ denoting the maximum inequality, where only one person possesses all the resources. 

There are two mathematical expressions of the Gini coefficient. For the first one, it is used to measure the variability of any statistical distribution. As proposed by Gini \cite{gini1912variabilita}, it is represented as the average absolute difference between pairs of observations $x,y$ divided by twice the mean $\mu$ , i.e.,
\begin{equation}
    G = \mathrm{E}(|x-y|)/(2\mu).
\end{equation}

\begin{figure}[htbp]
    \centering
    \includegraphics[width=0.5\textwidth]{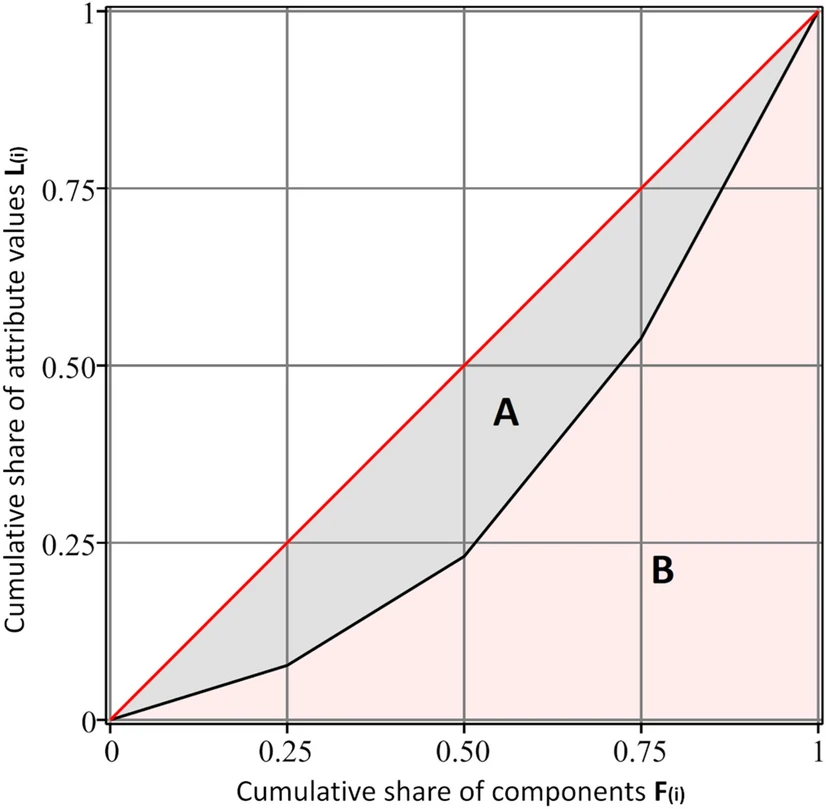}
    \caption{The Lorenz Curve for the Gini Coefficient \cite{lechthaler2020objective}.}
    \label{fig:gini}
\end{figure}

For the second one, it can be regarded as a summary statistic of the Lorenz Curve for a resource \cite{dorfman1979formula}. More specifically, as shown in Figure~\ref{fig:gini}, in which the $x$ axis represents the cumulative share of components $F(i)$ and the $y$ axis represents the cumulative share of attribute value $L(i)$, where $i$ is the index of population group, the Lorenz curve specifies the cumulative amount of the resource that belongs to the lowest $100i$-th percentile of the population. If it is perfect equality, the Lorenz curve converges towards the diagonal line (red line in the figure) from the origin $(0,0)$ to the upper right corner $(1,1)$, corresponding to a uniform distribution, which indicates that everyone receives the same amount of the resource. The general Lorenz curve can be represented as the black line in the figure, corresponding to the actual distribution of the resource. This black line splits the lower triangle into two areas $A$ and $B$, in which $A$ is the area between the egalitarian line and the general Lorenz curve, representing the gap between the actual allocation and the uniform distribution, and $B$ is the area below the Lorenz curve. Based on this, the Gini coefficient is defined as the ratio of $A$ to $A+B$, i.e., 
\begin{equation}
    G = A/(A+B).
    \label{eq:lo}
\end{equation}
Because $A+B=1/2$ and it is easy to obtain $B$, we can rewrite the above formula \eqref{eq:lo} for the discrete case as 
\begin{equation}
    G = 2A = 1 - 2B = 1 - \sum_{i=1}^N (F_i - F_{i-1})(L_i + L_{i-1}),
    \label{eq:gini}
\end{equation}
where $F_i$ is the cumulative proportion of the population up to group $i$ (with $F_0=0$), in which the population groups are non-decreasingly ordered, and $L_i$ is the cumulative proportion of the resource possessed by population groups up to group $i$ (with $L_0=0$). In the less-used continuous case, area $B$ can be measured by integration. Because the area under the equality line is constant, the further the Lorenz curve is away from the diagonal line, the higher the Gini coefficient gets, and the more severe inequality underlies. Yitzhaki and Schechtman \cite{yitzhaki2013gini} pointed out that the second definition can be applied to both discrete and continuous distributions while the first definition is only valid for the continuous case. Furthermore, different shapes of Lorenz curves may result in the same Gini coefficient value. In this case, the Gini coefficient ignores the structural difference among various inequalities. Therefore, when comparing similar Gini coefficients, a graphical representation based on the Lorenz curve is more precise in explaining the inequality structure.

\subsubsection{Application of Gini Coefficient in Energy Equity}


To quantify the equity issues in energy production, access, and consumption among different population groups, an increasing number of researchers have turned to Gini coefficients, as listed below. The Gini coefficient has been used to identify the distribution of energy among population groups \cite{SABOOHI2001245} before the concept of energy justice was involved. 

The first notable literature on Gini coefficient application in the context of energy justice focused on power consumption inequality \cite{jacobson2005letting}. In this work, the authors revealed the significant disparities across nations, using the intra-country distribution of residential electricity consumption with the Lorenz curves for five countries. By the comparison of Gini coefficients, they highlighted the stark contrast between Norway, where $38\%$ of the population used half of the residential electricity (Gini coefficient of 0.61), and Kenya, where $6\%$ of the households consumed half of its electricity (Gini coefficient of 0.87). Following this landmark study, many papers have employed the Gini coefficient to assess the energy consumption inequality. 

Some publications utilizing the Gini Coefficient of energy consumption focused on one country. In Brazil, the decreasing of the Gini Coefficient after the expansion of access to electricity endorses the effectiveness of rural electrification in terms of energy equity \cite{pereira2011challenge}. In China, fossil energy consumption was getting more equitable among provinces from 1997 to 2013 \cite{chen2017regional} while a reversal relationship between the Gini ratio and urbanization level indicates more energy equality among buildings in more developed provinces \cite{xiao2012reality}. In rural China, two major energy types contributing to the overall energy inequality were biomass ($61.4\%$) and electricity ($12.1\%$) \cite{wu2017measurement}. In both urban and rural India, equity in energy consumption was improved over 1987-2012 \cite{ranjan2017energy}. In the U.S., little inequality in residential energy consumption was seen during 1990-2015 while the energy poverty rate increased from $15.7\%$ in 1990 to $22.6\%$ in 2015 \cite{wang2021racial}.

Some publications took a global perspective. While global inequality in energy consumption has been decreasing \cite{lawrence2013global}, more prominent inequalities were evident across countries in specific energy services, including renewable energy \cite{duan2018analysis}, air transport energy \cite{oswald2020large}, and solid fuels \cite{rokicki2021changes}.

Beyond energy consumption, various Gini coefficients related to other energy equity have been explored, including carbon emission \cite{teng2011metric, chen2017decomposing, yang2019analysis}, energy services \cite{alstone2015decentralized}, photovoltaic solar installation \cite{tidwell2018surveying}, energy supply \cite{rokicki2020changes, rokicki2021diversity}, energy system's reliability \cite{heylen2019fairness, zhang2021will}, energy technology \cite{sueyoshi2021new}, and global energy trade \cite{li2022impacts}.

In addition to depicting the static status of energy equity at a time point, the Gini coefficient can be further integrated into the decision-making process for dynamic analyses of energy equity. This integration allows for monitoring changes in Gini coefficients related to a specific energy topic, thus demonstrating the positive impact on energy equity resulting from proposed frameworks or models. For instance, one study adopted the efficiency-fairness trade-off curve for evaluating the electricity pricing scheme, where fairness is measured by the Gini Coefficient \cite{sun2011advances}. Another research compared the Gini coefficients of natural gas consumption for actual and simulated data to support the proposed tariff scheme over the existing one \cite{zeng2018price}. The Gini coefficient was also employed to measure the regional equity as the result of their decentralized renewable electricity generation allocation that balanced cost-efficiency and energy equity \cite{sasse2019distributional}. Humfrey et al. \cite{humfrey2019dynamic} monitored the resulting Gini coefficient of $\mathrm{CO_2}$ emission from the proposed dynamic charging system integrated with renewable energy to ensure a fair impact on user experience.

Moreover, the Gini coefficient can be set as part of the objective to be optimized for the consideration of energy equity. For instance, one study introduced a preference for equality measured by the Gini coefficient in their generation expansion planning model for people with limited electricity infrastructure to appraise the trade-offs between energy inequality and total electricity consumption \cite{nock2020changing}. Ma et al. \cite{ma2020efficiency} included minimizing the Gini coefficient of coal production capacity in their coal capacity allocation model to balance efficiency and equity. Sasse and Trutnevyte \cite{sasse2020regional} modeled the regional equality of the impacts of energy transition using the Gini coefficient and discussed the trade-offs among minimizing system costs, maximizing regional equality, and maximizing renewable electricity generation. Levinson and Silva \cite{levinson2022electric} constructed a cost redistribution model for electricity pricing based on the Gini coefficients of electricity bills for each electricity utility in the US which shifted more costs in utilities with higher Ginis.

Publications also pointed out caveats of the Gini method when applied in energy equity studies. Considering the financial origin of the Gini coefficient, Jacobson et al. \cite{jacobson2005letting} emphasized the difference between income and energy data when applying the Lorenz curve. The authors stated two requirements when quantifying energy as a reasonable attribute for the Gini coefficient: (i) roughly constant overall energy efficiency among consumers, and (ii) approximately consistent marginal utility from the energy for each consumer. In other words, all individuals should produce similar results with the same level of energy usage and all individuals should have similar improvement with one unit of extra energy. Otherwise, the Gini coefficient may lose reliability when an equal amount of energy generates distinct levels of utility for different users. Shu and Xiong \cite{shu2018gini} validated the continuous Gini coefficient method for the fat-tailed data of China's inter-regional energy consumption.



\ren{Now}, we summarize the energy equity applications of the Gini coefficient based on the attribute values and population group criteria and reveal weak points in current studies addressing the modern concept of energy equity.

\begin{figure}[htbp]
    \centering
    \includegraphics[width=0.5\textwidth]{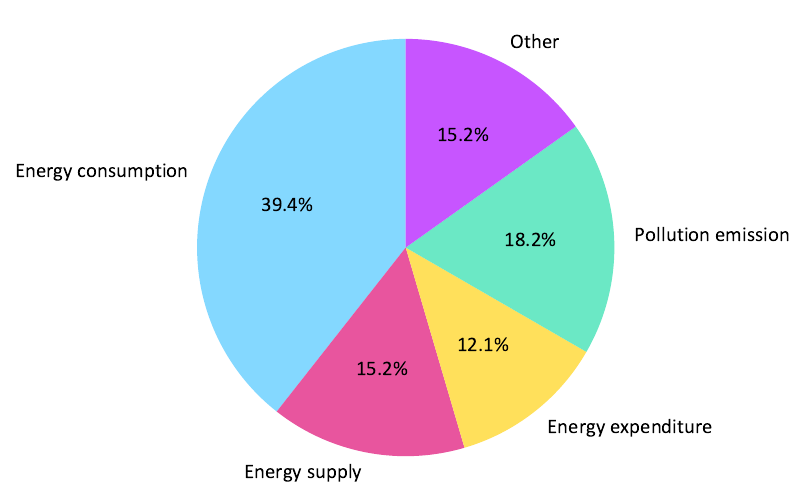}
    \caption{Attribute Values of Publications on Energy Equity with Gini Coefficient}
    \label{fig:gini-att}
\end{figure}

Table~\ref{tab:gini-att} and Figure~\ref{fig:gini-att} provide a summary of the attribute values used in the 33 publications discussed above. As can be seen, a large majority of the publications focused on energy consumption ($39\%$). Compared to other data, energy consumption data is more accessible, accurate, and relatively more stable concerning energy-based inequality \cite{wu2017measurement}. As shown in publications, inequality in energy consumption exists from emerging markets to developed areas, from past to present.

\begin{table}[htbp]
    \centering
    \caption{Publications on Energy Equity Using Gini Coefficient by Attribute Value}
    \begin{tabular}{p{0.25\linewidth}p{0.7\linewidth}} \toprule
    Attribute value & Publications \\ \midrule
    Energy consumption & \textcite{jacobson2005letting, pereira2011challenge, xiao2012reality, lawrence2013global, wu2017measurement, ranjan2017energy, chen2017regional, duan2018analysis, zeng2018price, oswald2020large, nock2020changing, rokicki2021changes, wang2021racial}\\
    Energy supply & \textcite{sasse2019distributional, ma2020efficiency, rokicki2020changes, rokicki2021diversity, li2022impacts}\\
    Energy expenditure & \textcite{alstone2015decentralized, ranjan2017energy, oswald2020large, levinson2022electric}\\
    Pollution emission & \textcite{teng2011metric, lawrence2013global, chen2017decomposing, shu2018gini, humfrey2019dynamic, yang2019analysis}\\
    Other & \textcite{sun2011advances, tidwell2018surveying, heylen2019fairness, zhang2021will, sueyoshi2021new}\\
    \bottomrule
    \end{tabular}
    \label{tab:gini-att}
\end{table}

However, consumption alone cannot depict the whole picture of the energy equity issues. Besides energy consumption revealing inequality in energy poverty, energy expenditure sheds light on energy accessibility and burden \cite{wang2021racial}, and energy production and supply affect energy security and diversification \cite{rokicki2020changes}. 

In addition to the general energy consumption data, researchers also have studied inequalities in terms of specific energy types and energy usages, revealing greater inequalities in less accessible energy sources (e.g., liquefied petroleum gas and biomass) and more expensive applications (e.g., space cooling and air transportation) \cite{wu2017measurement, oswald2020large}.

Most studies show energy inequality measured by direct data, while some recent publications utilize constructed indicators as attribute values for the Gini coefficient calculation. For example, Wang et al. \cite{wang2021racial} focused on the residential energy burden, defined as the ratio of the energy expenditure to the household income.

\begin{figure}[htbp]
    \centering
    \includegraphics[width=0.5\textwidth]{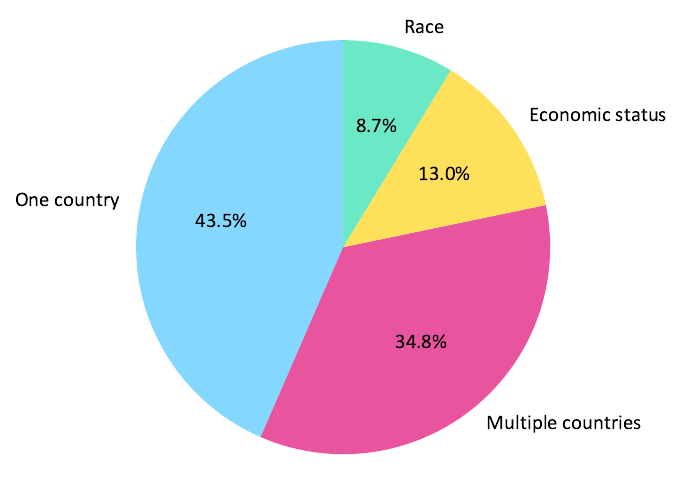}
    \caption{Comparison Scopes of Literature on Energy Equity with Gini Coefficient}
    \label{fig:gini-base}
\end{figure}

Finally, we report literature on energy equity based on the population group criterion. Table~\ref{tab:gini-base} and Figure~\ref{fig:gini-base} outline comparison scopes that have been published using the Gini method. Almost all Gini coefficients were calculated based on the population data grouped by region except for Zhang and Wang \cite{zhang2021will}, who used a GDP-based Gini. Hence, most of the publications focused on regional gaps ($78.3\%$), either concentrated on one country or investigated multiple countries at the same time. In the latter case, regional inequalities can be evaluated in two ways. Some studied global allocation of the energy by plotting multiple countries on the same Lorenz curve \cite{teng2011metric, tidwell2018surveying, duan2018analysis, mcgee2019renewable, oswald2020large, rokicki2021diversity, rokicki2021changes} while some displayed Lorenz curves for each country and compared the structural difference among their inequalities \cite{jacobson2005letting, ranjan2017energy, rokicki2020changes, tong2021measuring, zhang2021will}. 

\begin{table}[htbp]
    \centering
    \caption{Literature on Energy Equity using Gini Coefficient by Comparison Focus}
    \begin{tabular}{p{0.25\linewidth}p{0.7\linewidth}} \toprule
    Comparison focus  & Literature \\ \midrule
    Region & \\
    - One country & \textcite{pereira2011challenge, xiao2012reality, alstone2015decentralized, chen2017decomposing, sasse2019distributional, chen2017regional, shu2018gini, tidwell2018surveying, ma2020efficiency, zhang2021will}\\
    - Multiple countries & \textcite{jacobson2005letting, teng2011metric, lawrence2013global, duan2018analysis, mcgee2019renewable, rokicki2020changes, rokicki2021diversity, rokicki2021changes}\\
    Economic status & \textcite{ranjan2017energy, oswald2020large, zhang2021will}\\
    Race & \textcite{tong2021measuring, wang2021racial}\\ \bottomrule
    \end{tabular}
    \label{tab:gini-base}
\end{table}

Recent publications shifted their comparison scope from geography to other features, such as economic status and race. To evaluate disparities across prosperity levels, Zhang and Wang \cite{zhang2021will} established Lorenz curves based on GDP standards, and Ranjan and Singh \cite{ranjan2017energy} and Zhang and Wang \cite{zhang2021will} compared inequalities in rural and urban parts of a country. To study racial inequalities, Tong et al. \cite{tong2021measuring} represented each block of a city by its racial minority percentage, while Wang et al. \cite{wang2021racial} used household data categorized by ethnic group. There are two approaches to investigating inequalities among communities considering features, including region, race, gender, sexuality, and disability. One is using feature-based data, whose components can be divided into groups based on the characteristics of interest. The other one is using region-based data, where we can extract the feature for each region by quantifying its composition structure and distinguish the groups by regional difference. 

\subsubsection{Limitation of Gini Coefficient} \label{sec:gini-limit}

However, energy equality is not the same as energy justice. Despite wide applications, the Gini coefficient suffers from reductiveness that impedes the progress of solving energy injustices. It describes the state of inequality in a number while often ignoring the structure of the inequality which can be the center of today's energy issues. Hence, two very different Lorenz curves may produce the same value of the Gini index even though the actual distributions vary. For example, based on the data from the World Bank, in 2015, Greece and Thailand had the same Gini index of income ($0.36$) but the ratio of the income share held by the richest $10\%$ to the income share held by the poorest $10\%$ in Greece is $13.8$ while that in Thailand is $8.9$ \cite{sitthiyot2020simple}.

\begin{figure}[htbp]
    \centering
    \includegraphics[width=0.6\textwidth]{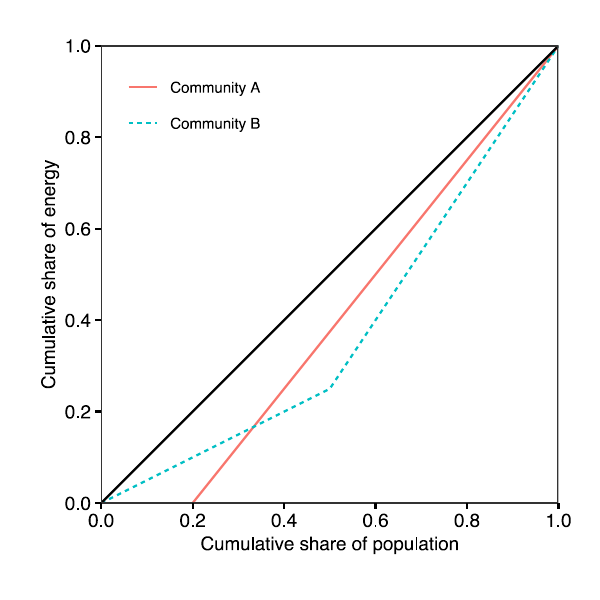}
    \caption{Lorenz Curve of Energy Distribution. The diagonal line denotes the perfect equality. The Gini coefficient of community A is $0.2$ and the Gini coefficient of community B is $0.25$.}
    \label{fig:gini-limitation}
\end{figure}

Furthermore, a situation with a certain amount of population inaccessible to the resource may still exhibit a good Gini coefficient as long as the resource is more equally distributed among the rest of the population albeit the people who suffer most are usually the focus of social problems. Graphically, this scenario can be represented as the lower end of a Lorenz curve staying at $0$ for some population shares. For example, consider two communities, A and B, with the same $100$ individuals who divide $100$ units of energy in total. Twenty people in community A are allocated $0$ energy resources while the rest of the community divides the energy equally. In community B, half of the population divides $25$ units of energy evenly while the other half consumes the rest $75$ units of the energy uniformly. As shown in Figure~\ref{fig:gini-limitation}, community A has a Gini coefficient of $0.2$ and community B has $0.25$. Because community A has a lower Gini value, we may conclude A is a better community regarding equity in energy consumption. However, there is a $20\%$ population that has no access to the energy in A while everyone in B has a similar amount of energy. The underserved people in community A are the focus of Executive Order 1398 signed by President Biden. Hence, this limitation makes the Gini coefficient ineffective in addressing the energy inequity issues for underserved communities. 

Moreover, the underserved community remains on the frontier that suffers from the fatal impacts of energy inequity. In reality, lack of access to energy services costs lives. An average of 702 heat-related deaths occurred in the US each year from 2004 to 2018 \cite{vaidyanathan2020heat}, the key driver of which is inaccessibility to the cooling service \cite{cardoza2020heat, nbc20210820}. In 2020, there were more than 20 million American families owed about $\$23$ billion in utility bills in total \cite{cnn20220428}. Therefore, failure to address the structural characteristics of the observed inequality when applying the Gini method can lead to ineffective or even false results.

\subsection{Generalized Entropy Indices}

\subsubsection{Generalized Entropy}
Besides the Gini Coefficient, inequality measures derived from the Generalized Entropy (GE) are another group of popular indices for evaluating income inequality. First used in thermodynamics, Entropy was introduced to information theory by Shannon as the information of a random variable \cite{amigo2018brief}. Generalized Entropy is defined as \cite{rohde2008lorenz}
\begin{equation}
    GE(\alpha) = \left[1/n \sum_{i=1}^n (q_i/p_i)^\alpha -1 \right] / (\alpha^2 - \alpha),
    \label{theil:ge}
\end{equation}
where $q_i = L_i - L_{i-1}$ denotes the resource share in group $i$ and $p_i = F_i - F_{i-1}$ denotes the population share in group $i$ with $L_i$ and $F_i$ defined the same as in \eqref{eq:gini}, and $\alpha$ is a non-negative sensitivity parameter. \ren{Note here $\alpha \ne 1$ and $GE(\alpha) \ge 0$ for all $\alpha \ne 1$. Further,} GE indices vary from $0$ to infinity with $0$ denoting the ideal egalitarian state of equal allocation and higher values meaning more severe inequality. The parameter $\alpha$ indicates the weight on the higher and lower ends of the distribution. Low values for $\alpha$ make GE more sensitive to changes in the lower tail, while higher values emphasize the population with more resources. 

\subsubsection{GE Indices}

Theil's $T$ \cite{theil1967economics} and Theil's $L$ \cite{theil1979world} are two commonly used Entropy indices \ren{with $p_i = 1/n$}. Taking the limits of \eqref{theil:ge} when $\alpha$ goes to $1$ or $0$ gives Theil's $T$ (see~\eqref{TheilT}) or $L$(see~\eqref{TheilL}) \ren{(the derivation is provided in Appendix)}, which are given by
\begin{equation}
    T = \sum_{i=1}^n q_i \ln (q_i/p_i),    \label{TheilT}
\end{equation}
\begin{equation}
    L = \sum_{i=1}^n p_i \ln (p_i/q_i).   \label{TheilL}
\end{equation}
Theil's $L$ sometimes is referred to as the mean log deviation measure. It is worth noting that the specification of $\alpha$ affects the configuration of the $GE(\alpha)$ index, making it hard for comparison between studies \cite{sitthiyot2020simple}.


While this class of equity measures is more complicated and graphically lacks intuitive representation, they help decompose inequalities for in-depth analysis, which is hard to accomplish with the Gini coefficient. If we arrange the groups into different categories, then inequalities measured by Theil indices can be decomposed into the ``within-category" component and the ``between-category" component. Depending on the categorizing criterion (e.g., region, GDP, education, or gender), decomposition can be helpful in policy-making and planning \cite{haughton2009handbook}.

Atkinson index is another \ren{equity index} related to GE. \ren{It} ranges from $0$ to $1$ (with $0$ representing ideal equality) and is defined as follows:
\begin{equation}
    A(\epsilon) =
    \begin{cases}
    1- \left[1/n \sum_{i=1}^n (q_i/\bar{q})^{1-\epsilon} \right]^{1/(1-\epsilon)}, & \epsilon \neq 1,\\
    1 - \left(\Pi_{i=1}^n q_i^{(1/n)} \right) / \bar{q}, & \epsilon = 1.
    \end{cases}    \label{eq:Atakin}
\end{equation}
where $\epsilon$ is an inequality aversion parameter and $\bar{q}$ is the average resource share per group. When $\epsilon$ increases to infinity, \ren{the smallest group resource share, i.e., $q_1 = \min_{i=1,...,n} q_i$, dominates the value of $A(\epsilon)$, because $q_1 < \bar{q}$. It is worth mentioning that the order sequences of the values obtained using the Atkinson index \eqref{eq:Atakin} and GE index \eqref{theil:ge} are the same} when $\alpha = 1 - \epsilon$ \cite{cowell1995measuring}.

\subsubsection{Application of GE Indices}

These GE indices have provided insights into equity issues within energy-related domains. Notably, three prominent areas of examination include energy intensity, energy consumption, and carbon emissions.

\ren{The GE index Theil's $L$ was first proposed to assess energy intensity (measured by the quantity of energy required per unit output or activity) inequalities across Organization for Economic Co-operation and Development (OECD) countries during 1971-1999 \cite{alcantara2004inequality}. In this study, the authors divided OECD countries into $4$ groups by geographical neighborhood and concluded that between-group inequality was the main contributor to the whole inequality. It further shows that the energy intensity inequality was alleviated during this study period, by reporting a 64\% decrease in the Theil index.} Another study applied Theil's $T$ on China's energy intensity data from 1985 to 2010 and illustrated narrowing gaps in regional inequality starting from 2004 \cite{feng2016weighted}.

\ren{For e}nergy consumption equity, Theil indices were utilized to analyze regional disparities of energy consumption based on 57 countries' data over the period 1995-2018 \cite{yao2020inequalities}. This study depicted a dynamic evolution of inequalities among geographical country groups for different energy sources, including oil, coal, natural gas, hydroelectricity, and renewable energy. Another study measured Theil indices of energy consumption in five countries of the Eurasian Economic Union from 2000-2017, revealing decreasing GDP-related inequality and increasing population-related inequality \cite{bianco2021energy}. In addition to measuring the energy consumption inequality among countries, researchers also used GE indices to reveal within-country disparities, such as in Indonesia \cite{dwi2020modern} and China \cite{ma2021energy}. Moreover, it is noteworthy that GE indices are often used complementary to the Gini Coefficient, demonstrating the reliability of the inequality result \cite{saez2018gross, takada2020gini, setyawan2020energy}.

\ren{F}or carbon emission equity, Theil's $L$ was used to investigate the evolution of $\mathrm{CO_2}$ emission inequalities in the EU during 1990-2009 \cite{padilla2013explanatory}, revealing a reduction in inequality between the groups of countries and providing insights for future environmental agreements. Later, for the EU from 2008-2016, a stable trend of carbon emission inequality was spotted using the GE indices \cite{bianco2019understanding}. Besides, research utilizing GE indices revealed an increasing trend of provincial inequalities of $\mathrm{CO_2}$ emission in Iran's agriculture sector \cite{pakrooh2020focus}.

\subsection{Other Equity Measures}
\ren{Besides Gini Coefficient and GE indices, there are several variance-based methods for} energy-related studies.

First, two variance-based statistics, standard deviation and coefficient of variation, have been applied as fairness indices to measure the inequality and inequity of the distribution of reliability among users in power systems \cite{heylen2019fairness}. Standard deviation, or the square root of the variance, measures the amount of dispersion of data points around their mean. A standard deviation of 0 indicates perfect equality. However, due to its lack of scale invariance, comparing inequality levels across diverse datasets can be challenging. In contrast, the coefficient of variation, defined as the ratio of standard deviation to the mean, offers a scale-adjusted assessment, aiding in the comparison of inequality across different contexts.

Next, Quality of Service (QoS) and Quality of Experience (QoE)\ren{, which are commonly used in telecommunications, have been adapted to guide equitable decision-making in energy} \cite{8301552}. QoS and QoE consider fairness based on service and user experience. Specifically, QoS measures the overall performance of a service which depends on the fulfillment of the user's need. Jain's index ($J$) is the most often used metric for QoS fairness \cite{hossfeld2016definition} and can be formulated as 
\begin{equation}
    J = 1 / (1 + c^2),
\end{equation}
where $c=\sigma / \mu$ is the coefficient of variation of QoS value. It ranges from $0$ to $1$ and reaches $1$ when the standard deviation is $0$, i.e., the same QoS across all entities. 

QoE measures ``the degree of delight or annoyance of the user of an application or service'' \cite{danner2021quality}. Fairness index $F$ is often used to evaluate the equity of QoE values among surveyed users. It is defined as
\begin{equation}
    F = 1 - \sigma / \sigma_{\max} = 1 - 2\sigma / (H- L),
\end{equation}
where $\sigma$ is observed standard deviation of the surveyed QoE values, $\sigma_{\max}$ is the maximum possible standard deviation. $\sigma_{\max}$ is calculated by the difference between $H$ and $L$, which are the upper bound and lower bound of the QoE value. By normalizing the standard deviation, index $F$ transforms the $\sigma$ of QoE to interval $[0,1]$ and attains the endpoint at minimum or maximum standard deviation.

\ren{For the usage} in various energy-related contexts, Jain's index was utilized along with the Gini coefficient as fairness indicators in an electricity pricing scheme, which aimed to maximize social welfare while enhancing fairness \cite{sun2011advances}. \ren{Both Jain's index and $F$ index are used in} \cite{8301552} to assess community fairness in their decentralized electricity market design, where prosumers (consumers who also produce power) share energy with the community. In \ren{the proposed} framework, QoS \ren{represents} the energy allocated to each individual, while QoE \ren{represents} the perceived price of energy. 

These fairness indices \ren{are also adapted to} electrical vehicles (EVs) \ren{charging} research. A fair charging mechanism for electrical vehicles, based on a queuing system, was proposed \ren{in} \cite{danner2021quality}. The study defined three QoS and two QoE criteria for the charging service and calculated their fairness with $F$ index. Through the integration of these metrics, their weighted fair queuing approach demonstrated superior performance in terms of QoS results and fairness level. \ren{Also,} transactive energy framework \ren{was established} for the real-time coordination of EVs with voltage control \cite{hoque2021transactive}. The fairness among EV owners {was measured} regarding energy trading volume (QoS) and their satisfaction (QoE) using Jain's and $F$ indices, respectively.


\section{\ren{Incorporating Energy Justice in Policy, Planning, and Operations}}

While defining, categorizing, and measuring energy justice is important, the true challenge lies in taking actionable steps to improve the current state of energy justice. Advancing energy justice can start from three primary aspects: implementing policies to assist disadvantaged communities, incorporating equity considerations into power system planning, and ensuring that energy operations are conducted with a focus on equity. In this section, we highlight recent endeavors aimed at promoting energy justice in policy, planning, and operations, while also delineating opportunities for future enhancements and progress.

\begin{figure}[htbp]
    \centering
    \includegraphics[width=0.8\textwidth]{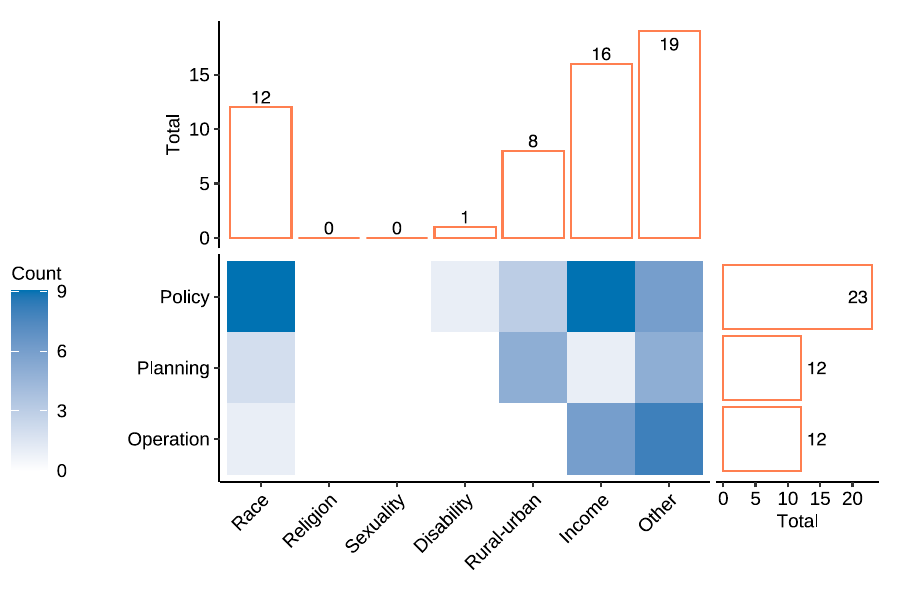}
    \caption{Heat map of publication count on integrating energy justice by category and equity concerns. Each literature is categorized into one or more tiles of the heat map based on its equity concern. The darker blue color of a tile means more publication while White tiles mean no publications identified by authors. Besides, the top bar plot displays the total count of unique publications for each equity concern; the right bar plot shows the total count of unique publications by category.}
    \label{fig:improve}
\end{figure}

Figure~\ref{fig:improve} provides a breakdown of publications discussed in this section, categorized by their integration category and equity concerns. Notably, previous research in policy advancement has predominantly highlighted disparities across races and income levels, while studies centered on planning have primarily focused on rural-urban inequalities. Emerging trends in equity concerns reveal an increasing interest in disabled communities \cite{darghouth2022characterizing}, while significant gaps related to religion or sexuality await attention.

\subsection{Policy}

Policies are guidelines or programs adopted by governments or other entities to guide decisions and achieve desired outcomes. Policies on improving energy justice have been focused on bill assistance, electricity pricing, and infrastructure deployment. However, these policies may face challenges of inefficient progress or unintended consequences.

Bill assistance policies can be categorized into two types based on the target population. The first type primarily concentrates on aiding low-income families, constituting the majority of bill assistance initiatives. Despite the implementation of various low-income energy programs like the Low-Income Home Energy Assistance Program (LIHEAP) in the US, energy poverty and insecurity persist among low-income households, particularly in rural areas, minority communities, and households with vulnerable members \cite{lewis2020energy}. Researchers criticized the short-term nature of these programs may be insufficient to address the enduring and structurally racialized consequences of energy insecurity \cite{brown2020high}.

The second type of bill assistance policy does not exclusively target the low-income population. Government and utility programs subsidizing residential rooftop solar power and EV adoption often remain out of reach for low-income households. While rooftop solar holds promise for transforming the US electric grid, high adoption costs have led to disparities in installation rates across demographics and regions \cite{darghouth2022characterizing}. To address these inequities, there is a growing focus on ensuring that low- and moderate-income (LMI) residents benefit from solar power. Initiatives like the State Energy Strategies project aim to expand the adoption of solar photovoltaic (PV) technology among LMI residents and communities, to benefit at least 15,000 LMI households \cite{leon2021state}. Similarly, policies that subsidize EVs to promote them as a clean alternative mode of transportation have also faced equity challenges. An equity analysis of the California Clean Vehicle Rebate Project discovered that, while the share of rebates distributed to low-income groups and disadvantaged communities increased over time, high-income buyers received the most share of rebates \cite{guo2021disparities}. They emphasized the need for more equitable incentive policies that take into account underserved individuals and communities.

Next, electricity pricing is another way to address energy justice through policy. It recovers utility costs for generation and indirectly affects energy justice compared to bill assistance \cite{ansarin2022review}. In general, the common equity considerations in electricity pricing can be summarized as allocative equity, distributional equity, and transitional equity \cite{burger2019fair}. Specifically, an allocative equitable tariff charges customers at the same location and time the same marginal rate for electricity consumption; distributional equity requires a tariff design to address the equitable access to electricity services for vulnerable customers, as identified by local standards; transitional equity considers the fair impact on customers when transitioning to new tariff structures. In practice, lifeline rates and energy affordability policies have been proposed to subsidize electricity consumption for low-income households \cite{iwuagwu2001electricity, nysix2019understanding}. Besides, a justice-aware tariff design has been introduced to account for socio-economic and demographic profiles to create an equitable locational hourly tariff \cite{khan2021electricity}. Moreover, while transitioning to more economically efficient tariffs has been observed to lead to bill increases for low-income customers in some cases, simple adjustments to tariff designs, such as discounted fix charges for low-income customers, can help mitigate these disparities \cite{burger2020efficiency}. Interestingly, a recent study suggests that bill assistance, such as providing direct financial compensation to low-income households, would be more economically efficient than subsidizing their utility rates \cite{chen2023optimal}.

In contrast, infrastructure deployment policies focus on improving energy efficiency and have a longer-term impact on alleviating energy inequity. Weatherization and community Distributed Energy Resources (DERs) emerge as focal points of such policies. The Weatherization Assistance Program (WAP) funds low-income households for weatherization, aiming to improve energy efficiency in homes \cite{brown2020high}. Proposed initiatives like on-bill financing policies with split incentives seek to amplify the benefits of weatherization for renters while optimizing energy efficiency for landlords \cite{bird2012policy}. Notably, the Biden-Harris Administration has taken steps to bolster federal programs targeting disadvantaged communities, including LIHEAP, WAP, and clean energy programs, intending to lower energy burdens, increase access to clean energy, and modernize the grid \cite{factsheet2022}. 

Moreover, community DER policies prioritize disadvantaged communities as sites for emerging energy technologies. DERs are small-scale electricity generation modules that can be integrated into the power grid, providing significant advantages to retailers and consumers by reducing cost and emission and increasing power reliability. Programs such as the California Energy Storage Initiative (SB 700) strive to integrate revenue-generating energy storage systems in low-income and underserved communities, ensuring equitable access to this technology's advantages \cite{tarekegne2021energy}. Similarly, initiatives like the Oakland Clean Energy Initiative (OCEI) aim to replace legacy fossil fuel power plants with clean energy storage deployment, thereby addressing historical disparities in energy access and environmental impacts \cite{mcnamara2022seeking}.

Despite the promising development of these energy policies, inequitable consequences may arise if not given enough scrutiny. Policies prioritizing economic stimulation may inadvertently overlook energy justice, as seen in the UK's fracking policy for shale gas, which shifted focus from community protection to pro-industry benefits, resulting in distributive injustices \cite{cotton2017fair}. To mitigate such issues, experts advocate for procedural justice through the energy justice framework, empowering communities in decision-making processes. Moreover, studies have highlighted disparities in funding allocation within certain clean energy programs, underscoring the importance of ensuring equitable distribution among diverse demographic groups \cite{zhou2019justice}. Additionally, research suggests that asset tests in bill assistance programs disproportionately affect the lowest-income families and increase administrative costs \cite{graff2019red}. Removing such tests could potentially expand program accessibility, yet as of 2024, their persistence varies across states, indicating the need for further evaluation and reform \footnote{https://liheapch.acf.hhs.gov/tables/assets.htm}.

\subsection{Planning}


Power system planning involves the design and construction of electricity networks. Publications integrating energy justice into power systems planning have primarily focused on rural electrification, renewable generator allocation, and EV charging station planning. 

First, the planning of rural electrification brings utility services to less developed areas and combats energy poverty. While the progress in rural electrification has often been sluggish \cite{CILLER2020109624}, researchers started to incorporate energy justice in their electrification planning models. For instance, an electrification planning model minimizing energy access inequality has been proposed to determine the location, type, capacity, and timing of power system infrastructure additions \cite{TROTTER2019288}. Specifically, the model minimizes the electrification inequality by reducing the maximum absolute discrepancy between urban and rural electrification rates. Their application on Uganda's national power system suggests that the equal access goal for 2040 can be achieved by increasing the system cost by only 3\%. Similarly, a study on power system expansion planning in sub-Saharan Africa developed a mathematical model that sought to ensure equal electrification for all urban and rural regions across the country \cite{musselman2022impact}. This research highlights the significant impact of varying the timeline for achieving full electrification on costs and generation capacity. Furthermore, another study proposed a generation expansion and transmission planning model minimizing per-capita energy consumption \ren{to achieve justice}\cite{nock2020changing}. By incorporating stakeholders' preference for equality, this study found that a strong preference for equality led to lower overall electricity consumption but higher electrification costs, due to increased investment in transmission infrastructure. Conversely, indifference to equality resulted in higher overall consumption levels in urban areas but lower electrification costs.

Second, planning on the equitable allocation of renewable power plants has surged as renewable energy technologies advance \cite{iea2021world}. A study in Germany models the allocation of different types of renewable power plants, with a focus on equal distribution of the negative environmental impacts \cite{drechsler2017efficient}. They highlight the conflict between the pursuit of equity and planning efficiency. An efficient planning approach for energy generation follows energy harvesting conditions, such as placing wind turbines in windy northern regions and photovoltaic power plants in sunny southern areas, whereas equity considerations advocate for a more even and mixed allocation of renewable power plants, which may undermine overall efficiency. Another renewable power planning study in Switzerland quantified this cost-efficiency trade-off by examining the balance between regional equity and the total generation cost \cite{sasse2019distributional}. They found that $50\%$ increase in equity led to an $18\%$ increase in cost, highlighting the challenges of achieving both equity and cost-efficiency simultaneously in generation planning.

Last but not least, the rising popularity of EVs raises equitable concerns regarding the planning of EV charging stations. For example, in California, disparities in access to charging facilities have been recognized, with data indicating that Black and Hispanic majority neighborhoods have approximately half less access than other neighborhoods \cite{HSU202159}. To address these disparities, a comprehensive four-stage framework has been proposed to integrate spatial equity into urban EV charging station planning \cite{li2022spatial}. By evaluating the spatial autocorrelation in neighborhoods, the framework can detect significant spatial patterns of an EV charging station distribution in terms of residential accessibility. Meanwhile, a machine learning approach has been developed to optimize the placement of EV charging stations while considering spatial disparities \cite{roy2022examining}. Their model predicts the future charging station density in Orange County, California, using the Random Forest algorithm, and optimizes the access equity by weighing various factors, including vehicle ownership rates, population density, the percentage of minority populations, education levels, and household income. 

\subsection{Operation}


Power system operations describe the process of decision-making from one day to minutes before the power delivery. In contrast to planning, power system operations have a more immediate and tangible impact on users. While studies on achieving energy justice through power system operations are still in their infancy, existing literature has focused on system reliability, EV charging mechanisms, and energy community operations.

Power system reliability is significantly influenced by decisions made in power system operations, raising equity concerns. A study \cite{heylen2019fairness} introduced fairness ratios to assess the distribution of reliability among end-users based on the energy demand and various reliability indicators, such as load curtailment, total interruption duration, and interruption cost. \ren{These indicators are evaluated through equity measures}, including the Gini coefficient and variance-based methods, to assess the impact of reliability on consumers. This approach enables the real-time evaluation of operational decisions in the power system, ensuring that the distribution of reliability aligns with energy justice principles.

\ren{Integrating energy justice into the operational aspect of EV charging has been studied. For instance,} study examined the fairness of the charging order for EVs \cite{hussain2022fairness}. They proposed an EV charging ranking mechanism, based on the demand urgency and fairness, where the fairness is quantified by Jain's index. This method outperformed existing charging operation rules by maximizing the number of satisfied EVs.

Recently, the concept of \ren{self-sustaining} energy community is gaining more popularity as it operates relatively independently from the grid and demonstrates significant opportunities to alleviate energy poverty, boost social inclusion, promote energy-saving measures, and transit to more sustainable and environmentally friendly energy sources \cite{CASALICCHIO2022118447, day2020equitable}. From the operational perspective, a study \cite{8301552} demonstrates a fair \ren{self-sustaining} energy community model that reflects the true preference of consumers while protecting their privacy. In this optimization model, the goals of the community manager, including preserving fairness, were realized by integrating energy importing and exporting, penalties, and geographical preferences into the optimization objective. Another study on \ren{self-sustaining} energy community system factorized a fairness index into their optimization model, allowing for equitable allocation of benefits and the prevention of penalties based on each member's contribution to the community \cite{CASALICCHIO2022118447}. Their approach compensates for individuals who may be suffering from disadvantages, ensuring a more consistent distribution of benefits to all users.

\section{Challenges and Opportunities}

\ren{Based on discussions in previous sections, we can observe} limiting factors that persist in the way to achieve energy justice. \ren{Meanwhile, we explore potential corresponding} future work. The challenges and outlooks \ren{can} appear in energy justice recognition, tools of measurement, and integration in policy, planning, and operations, explored below.

\subsection{Challenge in Energy Justice Recognition}

\ren{Blurred distinctions among diverse energy justice and equity concepts increase the likelihood of overlooking or congesting research areas, obstructing the recognition of injustices.} While the framework of energy justice is well established, some categories of energy equity problems have been ambiguous, such as the mixed-use of ``energy poverty" and ``energy insecurity" in some literature. To mitigate such confusion, we provided \ren{clear} definitions of each category and distinguished them by their extent of severity in Section~\ref{sec:def-lit-equity}. \ren{Even though}, these classifications \ren{actually should} be further \ren{clearly described} by energy types, energy services, and target demographics, which may enhance the recognition process for energy justice.

\ren{Another important issue is the imbalanced recognition of energy justice. This includes three aspects:} First, \ren{although} energy poverty and burden are becoming more and more addressed, energy insecurity and vulnerability deserve more attention because they complement the landscape of energy justice. Second, communities suffer the most from energy injustice\ren{, as compared to less suffered communities,} lack commensurate studies. For example, a review paper on rural electrification highlights that only $15\%$ of articles focus on countries with less than $50\%$ electrification rate \ren{(defined as the percentage of the population with access to electricity)} \cite{AKBAS2022111935}. Third, the majority of the literature has focused on disparities across geographical regions or income levels, often overlooking otherwise marginalized communities. For example, existing energy justice papers on planning mostly focused on the urban-rural gap, often ignoring the special needs of other underserved communities. \ren{All three of these imbalanced recognitions could lead to biased energy metrics and assistance programs.}

\subsection{Challenge in Measuring Energy Justice}

Measuring energy justice from various perspectives provides quantitative insights and facilitates actionable progress. However, the measurement of energy justice faces challenges from a lack of high-quality data and limitations of existing equality indices. 

First, lack of data not only leads to inaccurate quantification of energy inequity but also influences policy designs, as the availability of collected data significantly influences the design and implementation of energy assistance policies and programs \cite{leon2021state}. Especially with various innovative energy justice metrics being proposed \cite{cong2022unveiling, bednar2020recognition, nock2020changing, musselman2022impact}, acquirement of such data becomes crucial in promoting these metrics universally. Besides, it is worth noting that the current data collection processes might suffer from static time frames and financially skewed surveys, suggesting periodic evaluation for such datasets \cite{bednar2020recognition}.

Second, the limitations of existing tools for measuring energy justice need careful consideration. For instance, as we discussed in Section~\ref{sec:gini-limit}, while the Gini Coefficient is widely applied in measuring energy equities, it reduces the complex social issues to a number and may cause an improper representation of inequalities. 

\subsection{Challenge in Integration of Energy Justice in Planning and Operations}

\ren{Current models have made substantial progress in incorporating energy justice into planning and operational issues. However, for widespread acceptance among planners and operators, these models are expected to accommodate more practical power system constraints. Additionally, their objectives could expand from using a single index, such as an economic goal, to explore other objectives, such as reliability, environmental impacts, food security, and gender equality could also be taken into account. Balancing these factors with economic objectives poses a significant challenge.}

Another issue arises from the hardware limitations of power systems, as the \ren{hardware of power system components} could restrict assistance programs to advance energy equity, which can cause delayed participation for a community and further exaggerated disparity gaps relative to other communities that have gained access earlier \cite{sunter2019disparities}. Hence, \ren{synchronizing hardware upgrade of power system is desired to} bring timely and equitable access for underserved communities.

In conclusion, concerted efforts to overcome these obstacles are imperative to realize the promise of energy justice and foster a more equitable and sustainable energy future.

\appendix
\section{Appendix} \label{app:ge}
Here, we illustrate the derivation of Theil's $T$ \eqref{TheilT} and Theil's $L$ \eqref{TheilL} from General Entropy \eqref{theil:ge}. For a constant $c$, according to L'Hôpital's rule, we have
\begin{equation}
    \lim_{\alpha \to c} GE(\alpha) =
    \lim_{\alpha \to c} \frac{1/n \sum_{i=1}^n (q_i/p_i)^\alpha -1 }{\alpha^2 - \alpha} =
    \lim_{\alpha \to c} \frac{1/n \sum_{i=1}^n (q_i/p_i)^\alpha \ln(q_i/p_i)}{2\alpha - 1} =
    \frac{1}{n(2c-1)} \sum_{i=1}^n (\frac{q_i}{p_i})^c \ln \frac{q_i}{p_i}.
\end{equation}
With $p_i = 1/n$, we can get
\begin{equation}
    \text{Theil's } T
    = \lim_{\alpha \to 1} GE(\alpha)
    = \frac{1}{n} \sum_{i=1}^n \frac{q_i}{p_i} \ln \frac{q_i}{p_i}
    = \sum_{i=1}^n q_i \ln \frac{q_i}{p_i},
\end{equation}
and
\begin{equation}
    \text{Theil's } L
    = \lim_{\alpha \to 0} GE(\alpha)
    = -\frac{1}{n} \sum_{i=1}^n \ln \frac{q_i}{p_i}
    = \sum_{i=1}^n p_i \ln \frac{p_i}{q_i}.
\end{equation}